# Accurate collection of reasons for treatment discontinuation to better define estimands in clinical trials

Yongming Qu*

Eli Lilly and Company, Indianapolis, IN 46285, USA

Robin D. White

Eli Lilly and Company, Indianapolis, IN 46285, USA

Stephen J. Ruberg

Analytix Thinking, LCC, Indianapolis, IN 46236, USA

*Correspondence: Yongming Qu, Department of Statistics, Data, and Analytics, Eli Lilly and Company, Lilly Corporate Center, Indianapolis, IN 46285, USA. Email: qu_yongming@lilly.com

Running Head: Collection of reasons for treatment discontinuation

Word Count: 2217

**Abstract**

**Background:**

Reasons for treatment discontinuation are important not only to understand the benefit and risk profile of experimental treatments, but also to help choose appropriate strategies to handle intercurrent events in defining estimands. The current case report form (CRF) commonly in use mixes the underlying reasons for treatment discontinuation and who makes the decision for treatment discontinuation, often resulting in an inaccurate collection of reasons for treatment discontinuation.

**Methods and results:**

We systematically reviewed and analyzed treatment discontinuation data from nine phase 2 and phase 3 studies for insulin peglispro. A total of 857 participants with treatment discontinuation were included in the analysis. Our review suggested that, due to the vague multiple-choice options for treatment discontinuation present in the CRF, different reasons were sometimes recorded for the same underlying reason for treatment discontinuation. Based on our review and analysis, we suggest an intermediate solution and a more systematic way to improve the current CRF for treatment discontinuations.

**Conclusion:**

This research provides insight and directions on how to optimize the CRF for recording treatment discontinuation. Further work needs to be done to build the learning into Clinical Data Interchange Standards Consortium standards.

**Introduction**

Currently, randomized clinical trials remain the gold standard to evaluate the efficacy and safety of a new treatment. Randomization provides the basis for drawing statistically valid causal inferences for treatment effect. However, in phase 2 and 3 trials of reasonable size and duration, it is common that some participants will not adhere to their assigned study treatment (e.g., permanently discontinuing the study medication early), thus preventing observation of potential outcomes under the assigned treatments. Such deviations from the planned, randomized study treatment can create confusion regarding what treatment effect is to be estimated (estimand), how to do the estimation (estimator), and how to interpret the study result (estimate).

To address the impact of deviation from randomized treatments on the interpretation of clinical trial results, the ICH E9 (R1) addendum was released in 2020 to provide a framework for defining estimands and sensitivity analyses.[1] An estimand should have five components: treatment condition of interest, population of participants, variable (endpoint), handling of relevant intercurrent events (ICEs), and population-level summary. Per the ICH E9 (R1), ICEs are "events occurring after treatment initiation that affect either the interpretation or the existence of the measurements associated with the clinical question of interest". The ICH E9 (R1) addendum discussed five strategies for handling ICEs: hypothetical strategy, treatment policy strategy, composite strategy, while-on-treatment strategy, and principal stratum strategy.

The choice of strategies to handle ICEs depends on the study objectives and the underlying reasons for the ICEs. The ICH E9 (R1) addendum points out, "The question of what the values for the variable of interest would have been if rescue medication had not been available may be an important one. In contrast, the question of what the values for the variable of interest would have been under the hypothetical condition that subjects who discontinued treatment because of adverse drug reaction had in fact continued with treatment, might not be justifiable as being of clinical or regulatory interest." Therefore, the spirit of the ICH E9 (R1) addendum encourages the use of different strategies according to the underlying reasons of ICEs.

Treatment discontinuation is one type of commonly seen ICEs. Incorporating treatment discontinuation in efficacy analysis according to the reasons of treatment discontinuation has been discussed in previous research.[2-4] A mix of strategies for handling ICEs in defining estimands has been applied in clinical trial data analyses and discussed in literature.[5-8] Similarly, the assumptions and methods of imputing missing values should be based on the reasons for missingness and the ICEs (if leading to missing values).[8] For example, a patient who leaves a study (thus discontinuing the study treatment) due to moving to a new location may be considered differently from a patient who experiences an adverse event and discontinues their involvement in the study.

Additionally, data for the reasons for treatment discontinuation are important to understand the benefit and risk of a new treatment. For example, data on participants discontinuing the treatment due to lack of efficacy may give an indication of the new treatment's benefit (or lack of), while data on participants discontinuing the treatment due to adverse events may give an indication of the new treatment's risk (or lack of). Akacha et al. proposed the proportion of participants discontinuing treatment due to adverse events and the proportion of participants discontinuing treatment due to lack of efficacy are two important estimands in their tripartite estimands approach.[9] This approach was subsequently illustrated using real clinical data by Qu et al.[10-11]

Therefore, accurate collection of the reasons for ICEs is critical in defining estimands and constructing estimators. The ICH E9 (R1) addendum also emphasizes the importance of accurate collection of information regarding intercurrent events, "The additional granularity, identifying different ICEs, is necessary if different strategies are to be used. If the intercurrent event for which a strategy should be selected depends not only on, for example, failure to continue with treatment, but also on the reason, magnitude, or timing associated with that failure, this additional information should be defined and recorded accurately in the clinical trial." Based on our experience and interactions with regulatory agencies and industry peers, there have been no issues in collecting rescue medications; however, the accurate reason of treatment discontinuation was often not collected well. The available options following Clinical Data Interchange Standards Consortium (CDISC) standard in the most used case report forms (CRF) mix the reasons for treatment discontinuation (e.g., adverse events, lack of efficacy, etc.) and who makes the decision for treatment discontinuation (subject withdrawal, physician's decision, etc.)[12]; thus, it does not accurately reflect the underlying reasons for treatment discontinuation.

Surprisingly, CRF designs with such an apparent drawback are widely used by pharmaceutical industry and supported by CDISC.

We reviewed treatment discontinuation data for a set of completed phase 2 and phase 3 studies for insulin peglispro and manually regrouped the reasons according to the free-text comments. Through the exercise, we recommended new categories for the CRF for treatment discontinuation.

**Methods**

Insulin peglispro is a daily basal insulin investigated in clinical development from 2008 to 2014 by Eli Lilly and Company. Nine global phase 2 and phase 3 studies (duration of 12 to 78 weeks) in participants with type 1 and 2 diabetes were conducted.[13-21] The sponsor terminated the development of insulin peglispro based on the undesired benefit and risk profile.

In this analysis, we summarized the reasons for treatment discontinuation based on the original CRF categories using an integrated database of the nine studies. In addition, we reviewed the free-text comments entered by investigators for treatment discontinuation to identify the primary underlying reason for treatment discontinuation. It is possible that the free-text comments may suggest participants discontinue study treatment due to more than one reason, such as injection reaction and unsatisfactory efficacy; therefore, we used the following hierarchical priority to determine the primary reason for treatment discontinuation: "due to adverse events" > "due to lack of efficacy" > "due to other reasons". For example, if a patient discontinued treatment due to adverse events and lack of efficacy, the primary reason would be "due to adverse events".

**Results**

A total of 6215 participants who were assigned to treatments were included in the analysis: 857 (14%) participants discontinued study medication and the remaining participants completed their randomized, assigned treatment for the duration of the trial. Figure 1(a) summarizes the reasons for treatment discontinuation based on the original categories in the CRF. Among those with treatment prematurely discontinued, 20% of participants discontinued due to "adverse events", 3% due to "death", and 10% due to "protocol violation". For a large proportion of treatment discontinuations, the reason provided captured who decided the treatment discontinuation (rather than the reason for discontinuation): 40% discontinued due to "withdrawal by subject", 13% due to "physician decision", and 3% due to "sponsor decision" or "protocol required discontinuation". There were 11% of treatment discontinuations due to "lost to follow-up", in which sites could not gain additional information even with multiple attempts. Upon reviewing the comment fields, we found that sites entered various "reasons" for treatment discontinuation for the same scenario. For example, in the phase 3 studies, participants were required to discontinue the study medication if the level of triglycerides was greater than 500 mg/dL; however, reasons selected for participants with a level of triglycerides greater than 500 mg/dL included:

- ADVERSE EVENT (event ID = "hypertriglyceridemia")
- SPONSOR DECISION
- PHYSICIAN DECISION
- PROTOCOL REQUIRED DISCONTINUATION
- PROTOCOL VIOLATION

Technically, all options except "protocol violation" were correct answers although only the "adverse event" reason was to be the intended selection in the CRF. We agree that study teams could be more diligent in querying sites for better collection of underlying reasons for discontinuation, but this phenomenon indicates a fundamental flaw in the current CRF regarding treatment disposition: possible answers are confusing and not mutually exclusive, and selection options mix potential answers to two questions (what the underlying reason for treatment discontinuation is and who makes the decision for treatment discontinuation).

The summary of reasons for treatment discontinuation as originally recorded as well as based on newly grouped categories from a manual review of the free-text comments are shown in Figures 1(a) and 1(b), respectively. Notably, a higher proportion of treatment discontinuations were due to "adverse events" or "death" compared to the original CRF category (32% vs. 23%). The misclassification rates were relatively consistent between studies (data not shown). Additionally, approximately 15% of treatment discontinuations were clearly not related to efficacy and safety but due to reasons such as travel, relocation, personal reasons, family reasons, inclusion/exclusion criteria not met at baseline, site closure, and health or lifestyle status change.[5] Table 1 provides a summary of the number of treatment discontinuations in each category before and after reclassification.

**Recommendation and conclusion**

Fundamental changes to the collection of study treatment discontinuations may take some time as the industry recognizes this problem and works toward a new consensus and a resulting update to CDISC standards. Figure 2 illustrates an intermediate solution of using subcategories when the category of "withdrawal by subject" or "physician decision" is selected. This approach has been implemented in some clinical trials at Eli Lilly and Company.

The best solution moving forward may be to redesign the CRF to make the choice of answers less ambiguous. After our review, and consistent with the ICH E9 (R1) definition of ICEs, we believe that *all study medication discontinuation reasons* can be grouped into six primary categories: "Death", "Adverse Events", "Lack of Efficacy", "Sufficient/Excessive Efficacy", "Administrative", and "lost to follow up". The first three categories have obvious meanings. The fourth category, "Excess Efficacy", may apply in some situations in which a patient feels "cured" of their disease, believes they are no longer in need of treatment (e.g., anti-infectives, depression, or other psychiatric conditions), or the condition of the disease is overly corrected (e.g., too low blood glucose in anti-diabetes treatment). The fifth category is a catch-all for reasons not related to the safety or efficacy of the study treatment.

Figure 3 illustrates a potential future CRF for treatment discontinuation using the six primary categories noted above with further choices. Note the categories in Figure 3 will have to be harmonized with CDISC standards. Therefore, work across the industry and with the CDISC will be required to refine this proposal and update the CDISC standards as necessary. A PHUSE working group (Implementation of Estimands (ICH E9 (R1)) using Data Standards) is currently underway in addressing this issue.

In this article, we investigated the issues for the current practice in treatment discontinuation CRF, one area that we think can improve. Another type of ICEs, use of concomitant medications for unsatisfactory efficacy, was not discussed as there is generally no issue in accurately collecting such data. This research should also provide insight on collecting reasons for some less common ICEs such as treatment switch and treatment interruptions. In clinical trials, participants may discontinue the treatment first and then discontinue from the study later, or participants may discontinue the treatment and study simultaneously. In addition to collecting accurate reasons for treatment discontinuation, we should generally encourage participants to stay in the study and reduce the number of lost-to-follow-up participants. The reasons for the treatment and study discontinuations can be the same or different. For example, a patient discontinues the study and thus discontinues treatment due to site closure or study early termination. Another patient could decide to discontinue the study medication due to lack of efficacy and decline further study participation due to scheduling conflicts. Therefore, collecting the reasons for treatment discontinuation and study discontinuation separately is a good clinical practice. The learning and insight for treatment discontinuation discussed in this article can also be applied to study discontinuation, although study discontinuation may have different categories.

For all 9 studies, the sponsor conducted start-up meetings before the clinical trials started to educate site personnel on study design, study procedure, and data collection. It seems better education to site personnel and more careful data monitoring during the course of the clinical trials would probably reduce the magnitude of inaccurate collection of the reasons for treatment discontinuation. Therefore, the CRF design proposed in this article will need to be combined with site personnel training and data monitoring to optimize the data collection for treatment discontinuation.

There are two limitations in this research. Firstly, results were only based on studies in diabetes. Secondly, analyses were post hoc; the free-text comments did not always reveal the accurate reasons for treatment discontinuation.

In summary, we have discussed the inadequacy of the current CRF in collection of the reasons for treatment discontinuation. This inadequacy can hinder appropriate statistical analysis of a clinical trial, especially in light of the ICH E9 (R1) and its recommendations and strategies for handling ICEs. Through a systematic review and analysis of nine diabetes clinical trials, we have suggested six primary categories of reasons for treatment discontinuation and proposed a prototype of the improved new CRF. Future work is needed to further refine the list of categories and create more appropriate, meaningful, and useful CDISC standards.

**Acknowledgements**

We would like to thank Shanthi Sethuraman and Rong Liu for their useful discussions in this research and their scientific review of this article. We would also like to thank Antonia Baldo for his editorial review.

**Funding.** No funding was received for this research.

**Conflict of Interest.** Yongming Qu and Robin White are employees and shareholders of Eli Lilly and Company. Stephen Ruberg serves as consultant for several pharmaceutical companies.

**Clinical trials used in this article**: Clinicaltrials.gov numbers: NCT01027871 (Phase 2 for type 2 diabetes), NCT01049412 (Phase 2 for type 1 diabetes), NCT01481779 (IMAGINE 1 Study), NCT01435616 (IMAGINE 2 Study), NCT01454284 (IMAGINE 3 Study), NCT01468987 (IMAGINE 4 Study), NCT01582451 (IMAGINE 5 Study), NCT01790438 (IMAGINE 6 Study), NCT01792284 (IMAGINE 7 Study)

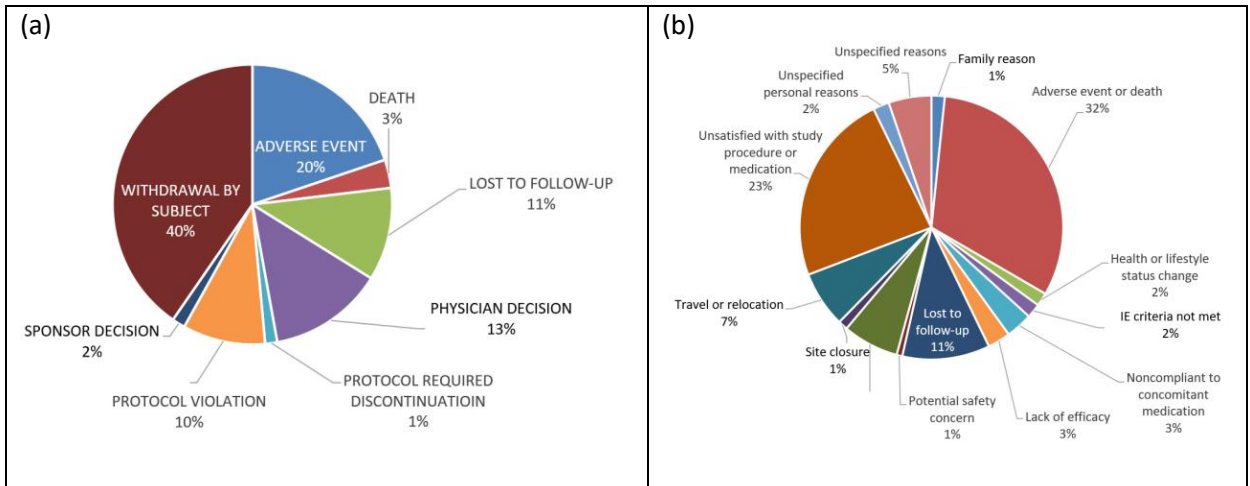

Figure 1. Summary of the reasons for treatment discontinuation by categories in the original CRF [Figure 1(a)] and by the newly grouped categories with post hoc review of the detailed comments [Figure 1(b)].

| What was the subject's status? | ○ Progressive Disease (PROGRESSIVE DISEASE) |
| | ○ Adverse Event (ADVERSE EVENT) |
| | ○ Death (DEATH) |
| | ○ Withdrawal by Subject (WITHDRAWAL BY SUBJECT) |
| | ○ Physician Decision (PHYSICIAN DECISION) |
| | ○ Non-Compliance With Study Drug (NON-COMPLIANCE WITH STUDY DRUG) |
| | ○ Protocol Deviation (PROTOCOL DEVIATION) |
| | ○ Study Terminated by IRB / ERB (STUDY TERMINATED BY IRB / ERB) |
| | ○ Study Terminated by Sponsor (STUDY TERMINATED BY SPONSOR) |
| | ○ Lost to follow up (LOST TO FOLLOW-UP) |
| | ○ Pregnancy (PREGNANCY) |

| Disposition Reason | Associated Sub-Categories |
|---|---|
| WITHDRAWAL BY SUBJECT | CONCERN ABOUT STUDY PROCEDURES/PERCEIVED RISKS |
| | HEALTH INSURANCE CHANGES |
| | SCHEDULING CONFLICTS |
| | SUBJECT IS MOVING OR HAS MOVED |
| | DUE TO EPIDEMIC/PANDEMIC |
| | PERSONAL ISSUE UNRELATED TO TRIAL |
| | OTHER (option to include a specify field) |
| PHYSICIAN DECISION | CONCERN ABOUT STUDY PROCEDURES/PERCEIVED RISKS |
| | HEALTH INSURANCE CHANGES |
| | SCHEDULING CONFLICTS |
| | SUBJECT IS MOVING OR HAS MOVED |
| | DUE TO EPIDEMIC/PANDEMIC |
| | OTHER (option to include a specify field) |

Figure 2. Temporary solution for better collection of the primary reason for treatment discontinuation

- DEATH
- ADVERSE EVENT. List the adverse event ID: _______________
- LACK OF EFFICACY
- SUFFICIENT/EXCESSIVE EFFICACY (if appropriate for the disease state under study)
- ADMINISTRATIVE (not related to safety and efficacy of study medication and

  Protocol Related
  - DID NOT MEET INCLUSION/EXCLUSION CRITERIA AT BASELINE
  - NONCOMPLIANCE TO STUDY MEDICATION
  - NONCOMPLIANCE TO STUDY PROCEDURE
  - NONCOMPLIANCE TO STUDY DRUG DELIVERY DEVICE/METHOD
  - UNSATISFIED WITH THE STUDY PROCEDURE
  - UNSATISFIED WITH THE STUDY DRUG DELIVERY DEVICE/METHOD
  - NEED TO TAKE PROTOCOL EXCLUDED CONCOMITANT MEDS

  Personal Circumstances
  - TRAVEL OR RELOCATION
  - SCHEDULE CONFLICT OR DIFFICULT TO TRAVEL TO SITES
  - UNSPECIFIED PERSONAL/FAMILY REASONS NOT RELATED TO EFFICACY OR SAFETY OF THE STUDY DRUG/DEVICE
  - UNEXPECTED EVENTS (NATURAL DISASTER, GEOGRAPHICAL CONFLICTS, OR PANDEMIC/EPIDEMIC)
  - PREGNANCY

  Study Logistics
  - STUDY TERMINATION, SITE CLOSURE, OR SITE PROCEDURE/SCHEDULE ERROR
  - DRUG SUPPLY CHAIN DISRUPTION
- LOST TO FOLLOW-UP

Figure 3. An illustration of the new CRF for the primary reason for treatment discontinuation

Table 1. Summary of the number of treatment discontinuations in each category before and after reclassification

| New Category | Original Category Collected from CRF | | | | | | | | | | |
|---|---|---|---|---|---|---|---|---|---|---|---|
| | ADVERSE EVENT | DEATH | INVESTIGATOR DECISION | LOST TO FOLLOW-UP | PHYSICIAN DECISION | PROTOCOL REQUIRED DISCONTINUATION | PROTOCOL VIOLATION | SPONSOR DECISION | SUBJECT DECISION | WITHDRAWAL BY SUBJECT | TOTAL |
| Adverse event or death | 170 | 28 | 3 | 0 | 22 | 8 | 4 | 4 | 2 | 31 | 272 |
| Family reason | 0 | 0 | 0 | 0 | 0 | 0 | 0 | 0 | 0 | 14 | 14 |
| Health or lifestyle status change | 0 | 0 | 0 | 0 | 3 | 1 | 2 | 1 | 0 | 7 | 14 |
| IE criteria not met | 0 | 0 | 0 | 0 | 0 | 0 | 15 | 0 | 0 | 0 | 15 |
| Noncompliant to concomitant medication | 0 | 0 | 0 | 0 | 1 | 1 | 22 | 1 | 0 | 3 | 28 |
| Lack of efficacy | 0 | 0 | 0 | 0 | 7 | 0 | 0 | 0 | 0 | 17 | 24 |
| Lost to follow-up | 0 | 0 | 0 | 92 | 0 | 0 | 0 | 0 | 0 | 0 | 92 |
| Potential safety concern | 0 | 0 | 0 | 0 | 1 | 0 | 0 | 0 | 0 | 5 | 6 |
| Schedule conflict | 0 | 0 | 0 | 0 | 1 | 0 | 0 | 0 | 0 | 58 | 59 |
| Site closure | 0 | 0 | 0 | 0 | 1 | 0 | 5 | 3 | 0 | 1 | 10 |
| Travel or relocation | 0 | 0 | 0 | 0 | 0 | 0 | 1 | 1 | 0 | 57 | 59 |
| Unsatisfied with study procedure or medication | 0 | 0 | 0 | 0 | 74 | 2 | 32 | 3 | 0 | 91 | 202 |
| Unspecified personal reasons | 0 | 0 | 0 | 0 | 0 | 0 | 0 | 0 | 0 | 17 | 17 |
| Unspecified reasons | 0 | 0 | 0 | 0 | 1 | 0 | 1 | 0 | 1 | 42 | 45 |
| Total | 170 | 28 | 3 | 92 | 111 | 12 | 82 | 13 | 3 | 343 | 857 |